\documentclass[pra,twocolumn]{revtex4-2}

\usepackage{epsfig,amsmath}
\usepackage{subfigure}
\usepackage{graphicx}
\usepackage{dcolumn}
\usepackage{stmaryrd}
\usepackage{mathrsfs}
\usepackage{pifont}
\usepackage{amsthm}
\usepackage{amssymb}
\usepackage{bm}
\usepackage{latexsym}
\usepackage{color}
\usepackage{hyperref}
\usepackage{multirow}
\usepackage{xcolor}
\usepackage{babel}
\usepackage[ruled]{algorithm2e}
\usepackage{verbatim}
\usepackage{enumerate}

\begin{document}
\title{Optimized control for high-fidelity state transmission in open systems}
\author{	
		Yang-Yang Xie\textsuperscript{1},
		Feng-Hua Ren\textsuperscript{2},
		Arapat Ablimit\textsuperscript{1},
		Xiang-Han Liang\textsuperscript{1},
		Zhao-Ming Wang\textsuperscript{1}
		}
\email{wangzhaoming@ouc.edu.cn}

\address{
		$^{1}$College of Physics and Optoelectronic Engineering, Ocean University of China, Qingdao 266100, People’s Republic of China\\
		$^{2}$ School of Information and Control Engineering, Qingdao University of Technology, Qingdao 266520, People’s Republic of China
		}
\date{\today}

\begin{abstract}
Quantum state transfer (QST) through spin chains has been extensively investigated. Two schemes, the coupling set for perfect state transfer (PST) or adding a leakage elimination operator (LEO) Hamiltonian have been proposed to boost the transmission fidelity. However, these ideal schemes are only suitable for closed systems and will lose their effectiveness in open ones. In this work, we invoke a well explored optimization algorithm, Adam, to expand the applicable range of PST couplings and LEO to the open systems. Our results show that although the transmission fidelity decreases with increasing system-bath coupling strength, Markovianity and  temperature for both ideal and optimized cases, the fidelities obtained by the optimized schemes always outweigh the ideal cases. The enhancement becomes more bigger for a stronger bath, indicating a stronger bath provides more space for the Adam to optimize. This method will be useful for the realization of high-fidelity information transfer in the presence of environment.
\end{abstract}
\maketitle

\section{Introduction}
High-fidelity information transfer between qubits lays a firm foundation for the realization of large-scale fault-tolerant quantum computers \cite{kimble2008quantum}. Spin qubits interact through nearest-neighbor Heisenberg exchange coupling and constitute a one-dimensional spin chain. Bose has proposed to use a spin chain as the channel for short-distance communication \cite{magnetic1}. Nonetheless, the transmission fidelity decreases with increasing number of spins \cite{magnetic1, hu2009state}. Lots of strategies for the fidelity improvement have been proposed, such as arranging the special couplings between nearest-neighbor sites for PST \cite{pst1, pst2}, adding well-designed external fields \cite{externalfield1, externalfield2, PhysRevA.82.022318, wang2020aest}. QST has also been experimentally investigated in varieties of platforms, including superconducting qubit chains \cite{li2018perfect}, trapped ions \cite{trappedions}, ultracold atoms \cite{ultracoldatoms}, semiconductor quantum dots \cite{quantumdots, kandel2021adiabatic}, etc.

Along with the above alluded ones, the proposed schemes are mainly based on ideally closed systems. When considering the environments \cite{arapat2022, wang2020aest2}, normally the information processing which can be performed well in closed systems will be destroyed by the system-bath interaction. The detrimental effects of a Markovian \cite{hu2009state} or non-Markovian \cite{environment1, nie2021, wang2020aest2} bath on the QST through spin chains have been investigated recently. The transmission fidelity is found to decrease with the increasing system-environment coupling strength, environmental characteristic frequency and temperature \cite{wang2020aest2, nie2021}. A lot of schemes have been proposed to reduce these adverse effects, like modulating the couplings between the spins \cite{pst1, pst2,hu2009state} or invoking an LEO \cite{wang2020aest2, nie2021}. In our recent work, we investigate the almost exact state transmission in a spin chain by adding an LEO Hamiltonian \cite{wang2020aest, wang2020aest2}. The LEO Hamiltonian can be realized by a sequence of control pulses. The pulse conditions have been obtained in a closed system for almost exact QST \cite{wang2020aest2}. When applying these conditions to an open system, the fidelity decreases due to the existence of the environments \cite{wang2020aest2, pulsecondition}.

Gradient descent is the most basic optimization algorithm \cite{zhang2019does}, moving relevant parameters towards the direction minimizing a predefined cost, or loss, function but without guaranteeing a fast and stable convergence. Momentum algorithm makes progress with this problem by updating parameters according to the gradients of current and previous iterations \cite{sutskever2013importance}. Besides, one of the algorithms with adaptive learning rates, RMSprop \cite{kurbiel2017training}, can modulate the learning rate on the basis of different parameters and training phases. Adaptive Moment Estimation (Adam) algorithm builds on and hence inherits the above two ones, realizing more efficient convergence behaviors, and become the most popular optimizer even in the noisy intermediate-scale quantum (NISQ) device era \cite{kingma2014adam, reddi2019convergence, he2023modularized}. Recently we use the stochastic gradient descent or Adam algorithm to find the optimized pulses for the adiabatic speedup \cite{PhysRevA.106.062612} or non-adiabatic QST \cite{liang2023optimally} in a noisy environment. The control pulses are designed via optimization algorithms by considering both the system and environment. As stated above, the ideal pulses are not effective for the open systems. In this paper, we use Adam algorithm to design the optimized couplings or pulses for high-fidelity QST through a spin chain in a non-Markovian environment. By defining an effective loss function which is relevant to the system and environmental parameters, the real unknown parameters can be revealed and the optimized solution is obtained along the gradient descent direction. We adopt a new-developed non-Markovian quantum master equation approach to solve the corresponding dynamics of the system \cite{wang2021stfinit}. For the optimized couplings, we find that the achievable maximum fidelity can be enhanced and the corresponding arrival time can be shortened as well. For the optimized control pulses, our results show that they can acquire better QST qualities than the ideal closed-system pulses do. In both scenarios, the effects of system-bath coupling strength $\Gamma$, environmental non-Markovianty $\gamma$ and temperature $T$ on the fidelity are analyzed. The fidelity decreases with increasing anyone of above parameters as expected, but the fidelity can be improved by our optimized schemes, especially in a strong environment.

\section{Model and method}

\subsection{The model and the Hamiltonian}
When a quantum system is exposed to its environment, the total Hamiltonian $H_{tot}$ consists of three parts
\begin{equation}
	H_{tot}=H_{s}+H_{b}+H_{int}.
\end{equation}
Here $H_{s}$ and $H_{b}=\sum_{k}\omega_{k}b_{k}^{\dag}b_{k}$ are the system and bath Hamiltonian, respectively. $H_{int}=\sum\limits_{k}(g_{k}^{\ast}L^{\dag}b_{k}+g_{k}Lb_{k}^{\dag})$ accounts for the interaction between them. $\omega_{k}$ indicates the $k$th mode frequency of bath and $b_{k}^{\dag}$ ($b_{k}$) represents the bosonic creation (annihilation) operator. The system is linearly coupled to a bosonic bath through the Lindblad operator $L$ with coupling constant $g_{k}$.

According to the QSD approach \cite{qsd1, qsd2, qsd3, wang2021stfinit}, the dynamical evolution of an open system in a non-Markovian finite-temperature heat bath is governed by
\begin{eqnarray}
	\frac{\partial}{\partial t}\rho_{s}&=&-i[H_{s},\rho_{s}]+[L,\rho_{s}\overline{O}_{z}^{\dag}(t)]-[L^{\dag},\overline{O}_{z}(t)\rho_{s}]\notag \\
	&&+[L^{\dag},\rho_{s}\overline{O}_{w}^{\dag}(t)]-[L,\overline{O}_{w}(t)\rho_{s}].
	\label{master}
\end{eqnarray}
The operators ${O}_{z(w)}$ are defined by an ansatz, and enter this evolution equation through the memory kernels $\overline{O}_{z(w)}(t)=\int_{0}^{t}ds{\alpha}_{z(w)}(t-s){O}_{z(w)}(t,s)$. To simplify, we adopt the weak system-bath coupling and low frequency (or high temperature) approximations. Moreover, the chosen spectrum density, Ohmic type with a Lorentz-Drude cutoff function \cite{ld1, ld2, ld3}, reads $J(\omega)=\frac{\Gamma}{\pi}\frac{\omega}{1+(\frac{\omega }{\gamma})^{2}}$. Subsequently, the two bath correlation functions ${\alpha}_{z(w)}(t-s)$ in $\overline{O}_{z(w)}(t)$ satisfy the following condition
\begin{equation}
	\frac{\partial\alpha_{z(w)}(t-s)}{\partial t}=-\gamma\alpha_{z(w)}(t-s).
\end{equation}
Then the operators $\overline{O}_{z,(w)}(t)$ obey the closed equations \cite{wang2021stfinit}
\begin{equation}
	\frac{\partial \overline{O}_{z}}{\partial t}=(\frac{\Gamma T\gamma}{2}-\frac{i\Gamma\gamma^{2}}{2})L-\gamma\overline{O}_{z}+[-iH_{s}-(L^{\dag }\overline{O}_{z}+L\overline{O}_{w}),\overline{O}_{z}], 
	\label{o1}
\end{equation}
\begin{equation}
	\frac{\partial \overline{O}_{w}}{\partial t}=\frac{\Gamma T\gamma}{2}L^{\dag}-\gamma\overline{O}_{w}+[-iH_{s}-(L^{\dag }\overline{O}_{z}+L\overline{O}_{w}),\overline{O}_{w}]. 
	\label{o2}
\end{equation}
As a result, we are allowed to numerically solve the dynamical evolution equation in Eq.~(\ref{master}), with the help of Eqs.~(\ref{o1}) and (\ref{o2}). In the above derivation, $\Gamma$ and $\gamma$ stand for the system-bath coupling strength and the characteristic frequency of bath. The Ornstein-Uhlenbeck correlation function $\Lambda(t,s)=\frac{\gamma}{2}e^{-\gamma\left\vert t-s\right\vert}$ contained in ${\alpha}_{z(w)}(t-s)$, decays exponentially with the environmental memory time $1/\gamma$ characterizing the memory capacity of the bath. Therefore for small $\gamma$, non-Markovian properties can be observed. The large $\gamma$ corresponds to a Markovian bath due to the shrinking environmental memory time.

When $\gamma$ approaches $\infty$, the bath becomes completely Markovian and has no memory capacity anymore. Consequently, $\overline{O}_{z}=\frac{\Gamma T}{2}L$ and $\overline{O}_{w}=\frac{\Gamma T}{2}L^{\dagger}$. The master equation in Eq.~(\ref{master}) therefore reduces to the Lindblad form \cite{wang2021stfinit}
\begin{eqnarray}
	\frac{\partial }{\partial t}\rho_{s}&=&-i[H_{s},\rho_{s}]+\frac{\Gamma T}{2}[(2L\rho_{s}L^{\dagger}-L^{\dagger}L\rho_{s}-\rho_{s}L^{\dagger}L)\notag \\
	&&+(2L^{\dag}\rho_{s}L-LL^{\dag}\rho_{s}-\rho_{s}LL^{\dag})].
	\label{markov}
\end{eqnarray}

In this paper, we consider a one-dimensional $XY$ spin chain as the system
\begin{equation}
	H_{s}=\sum_{i=1}^{N-1}J_{i,i+1}\left( \sigma_{i}^{x}\sigma_{i+1}^{x}+ \sigma_{i}^{y}\sigma_{i+1}^{y}\right ).
	\label{hs}
\end{equation}
Here $\sigma_{i}^{\alpha}$ ($\alpha=x, y$) stands for the Pauli operator acting on the $i$th spin. $J_{i,i+1}$ indicates the relevant coupling strength between the nearest-neighbor sites $i$, $i+1$ and we set the PST coupling layout $J_{i,i+1}=-\sqrt{i\left ( N-i\right )}$ throughout.

Initially prepare all the spins at the down state but the first one at the up state, i.e., $|\Psi_{s}\left(0\right)\rangle=|\textbf{1}\rangle=|100\dots0\rangle$. Our task is to transfer the state $|1\rangle$ from the first to the last spin of the chain, and the target state will be $|\textbf{N}\rangle=|000\dots1\rangle$. During this process, the transmission fidelity $F(t)=\sqrt{\left\langle \textbf{N}\right\vert \rho _{s}(t)\left\vert \textbf{N}\right\rangle}$ is monitored to evaluate the transfer quality. Here $\rho_{s}(t)$ is the reduced density matrix of our system.

Combining the advantages of two algorithms with fast and steady convergence, Momentum and RMSprop, Adam has already become the most valuable optimizer in the NISQ era. Now in this work we use the Adam to construct an iterative process to optimize the parameters for high-fidelity state transfer in noisy environments.

First we need to define a loss function $Loss$ and our goal of high-fidelity QST is encoded as to minimize the $Loss$. The specific optimization procedure of Adam algorithm is as follows.
\begin{enumerate}[\textit{Step 1:}]
	\item Compute the gradient vector $\textbf{\textit{g}}$ of loss function $Loss$ with respect to selected variables $\textbf{\textit{A}}$ in the $k$th iteration
	\begin{equation}
		\textbf{\textit{g}}^{k}=\triangledown_{\textbf{\textit{A}}^{k}}Loss(\textbf{\textit{A}}^{k}).
	\end{equation}
\end{enumerate}
\begin{enumerate}[\textit{Step 2:}]
	\item Compute the new exponential moving averages
	\begin{equation} 						
		\textbf{\textit{m}}^{k}=\beta_{1}\textbf{\textit{m}}^{k-1}+(1-\beta_{1})\textbf{\textit{g}}^{k},
	\end{equation}
	\begin{equation}
		\textbf{\textit{v}}^{k}=\beta_{2}\textbf{\textit{v}}^{k-1}+(1-\beta_{2})(\textbf{\textit{g}}^{k})^{2}.
	\end{equation}
\end{enumerate}
\begin{enumerate}[\textit{Step 3:}]
	\item Compute the new bias-corrected moment vectors
	\begin{equation} 							  
		\hat{\textbf{\textit{m}}^{k}}=\textbf{\textit{m}}^{k}/[1-(\beta_{1})^{k}],
	\end{equation}
	\begin{equation}
		\hat{\textbf{\textit{v}}^{k}}=\textbf{\textit{v}}^{k}/[1-(\beta_{2})^{k}].
	\end{equation}
\end{enumerate}
\begin{enumerate}[\textit{Step 4:}]
	\item Update the variables $\textbf{\textit{A}}$ according to
	\begin{equation} 							  
		\textbf{\textit{A}}^{k+1}=\textbf{\textit{A}}^{k}-\alpha \hat{\textbf{\textit{m}}^{k}}/(\sqrt{\hat{\textbf{\textit{v}}^{k}}}+\varepsilon).
	\end{equation}
\end{enumerate}
\begin{enumerate}[\textit{Step 5:}]
	\item Repeat steps $1-4$ till $Loss<\xi$ or the number of iterations $k>k_{max}$. $\xi$ (set $\xi =0.001$) and $k_{max}$ are the prescribed loss ceiling and maximal iteration number, respectively.
\end{enumerate}

\section{Results and discussions}

\begin{figure}[!htbp]
	\centering
	\includegraphics[width=0.95\columnwidth]{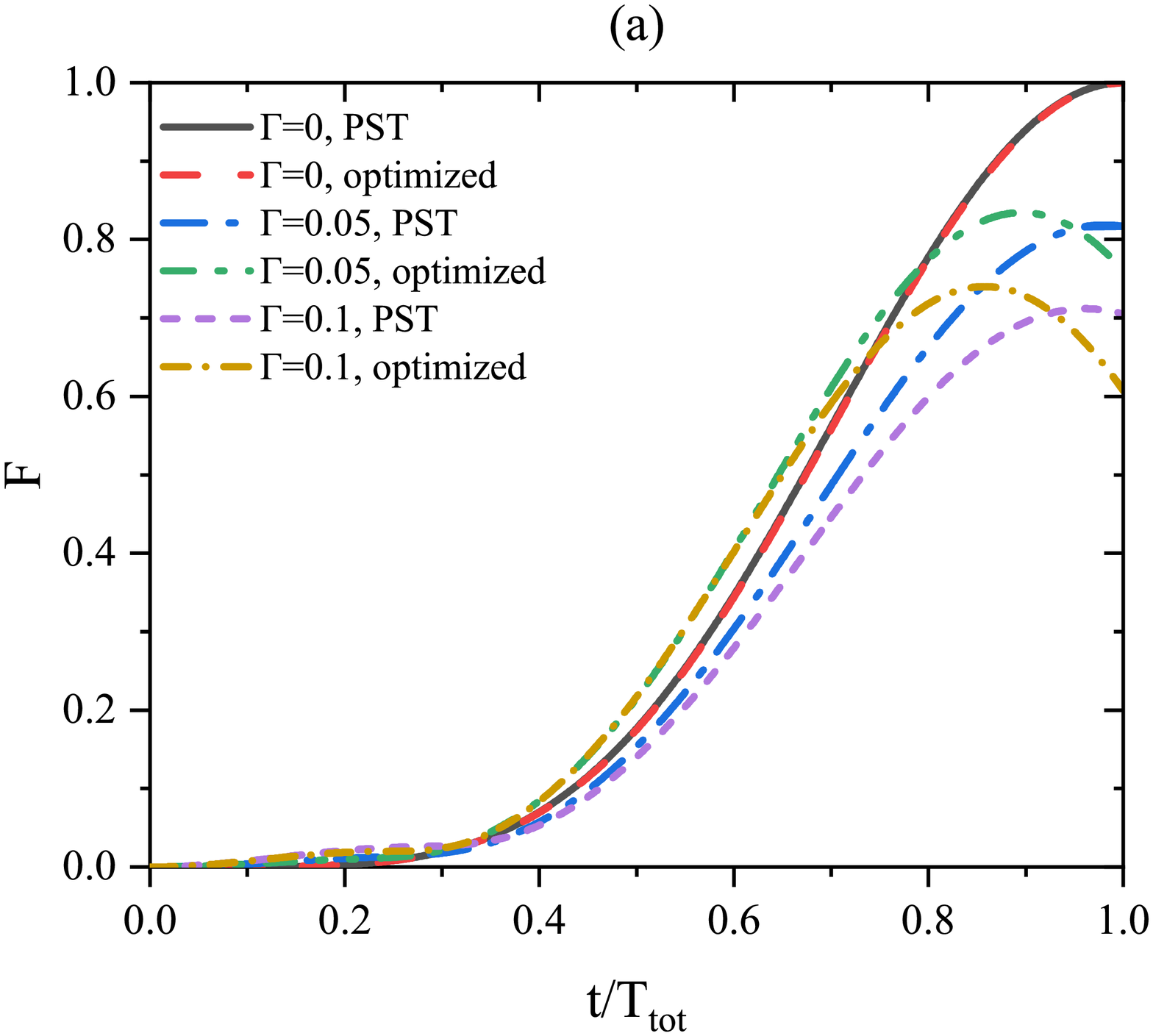}
	\includegraphics[width=0.95\columnwidth]{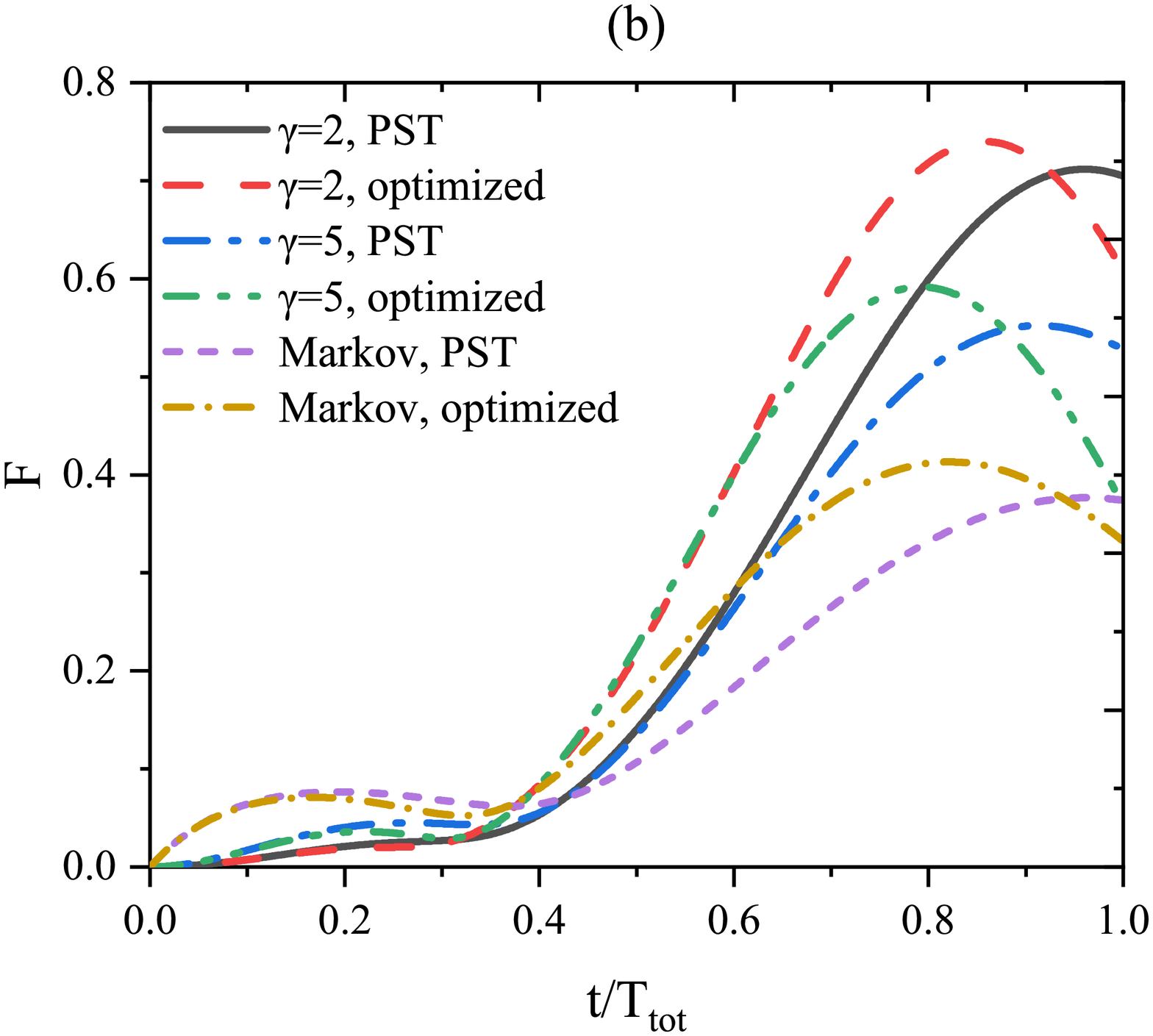}
	\includegraphics[width=0.95\columnwidth]{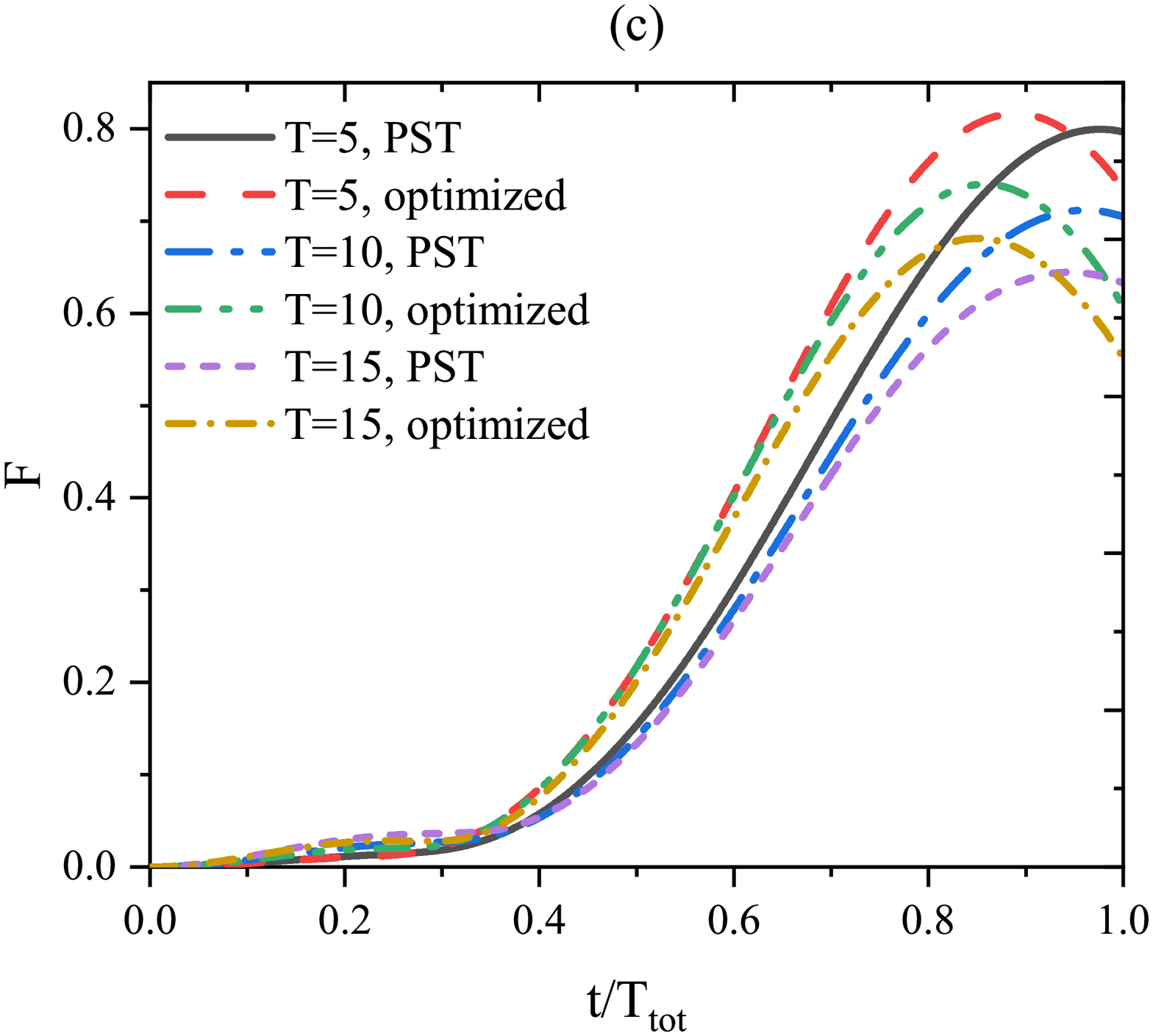}
	\caption{(Color on line) The fidelity $F$ versus the rescaled time $t/T_{tot}$ with PST and optimized couplings for different parameters (a) $\Gamma$, $\gamma=2$, $T=10$; (b) $\gamma$, $\Gamma=0.1$, $T=10$; (c) $T$, $\Gamma=0.1$, $\gamma=2$. $N=6$, $L=\sum_{i=1}^{N}\sigma _{i}^{-}$.}
	\label{f1}
\end{figure}

In this work, we consider two scenarios and apply Adam optimizer to explore high-fidelity QST through a spin chain in open systems. For the first one, we choose to modulate the coupling strength sequence $\textbf{\textit{J}}=[J_{1,2}, J_{2,3}, \cdots, J_{N-1,N}]$. For the second one, we optimize the pulse amplitude sequence $\textbf{\textit{I}}=[I_{0}, I_{1}, \cdots, I_{M-1}]$ to realize a more effective LEO. Without loss of generality, we consider the Lindblad operator $L=\sum_{i=1}^{N}\sigma _{i}^{-}$ that describes the dissipation. Here $\sigma _{i}^{-}=\left (\sigma _{i}^{x}-i\sigma _{i}^{y}  \right )/2$ denotes the lowering operator on the $i$th spin.

\subsection{Optimized couplings via Adam}
In this section, we perform the coupling optimization. Recall that our goal is to minimize a commonly defined loss function
\begin{equation}
	Loss(\textbf{\textit{J}})=1-F(\textbf{\textit{J}})+\lambda J_{max},
	\label{loss1}
\end{equation}
where the fidelity $F(\textbf{\textit{J}})$ is obtained with the help of the optimized coupling sequence $\textbf{\textit{J}}$ and $Loss(\textbf{\textit{J}})$ is the corresponding infidelity. $J_{max}$ stands for the maximal absolute value of couplings $J_{i,i+1}$ in optimized couplings. The relaxation parameter $\lambda$ is introduced here to modulate the proportion of $J_{max}$ in $Loss$ to restrain $J_{max}$ to not too large.

As an example, the number of spins is taken as $N=6$. Here we take the PST couplings $J_{i,i+1}=-\sqrt{i\left ( N-i\right )}$ as an initial guess and set the maximal iteration number $k_{max}=1000$. In addition, it is necessary to mention that in closed systems, PST can be observed at $t=n\pi /4$ ($n$ is an odd integer) for the PST couplings. Accordingly, the total evolution time is taken as $T_{tot}=\pi /4$ throughout.

\begin{figure}[!htbp]
	\centering
	\includegraphics[width=\columnwidth]{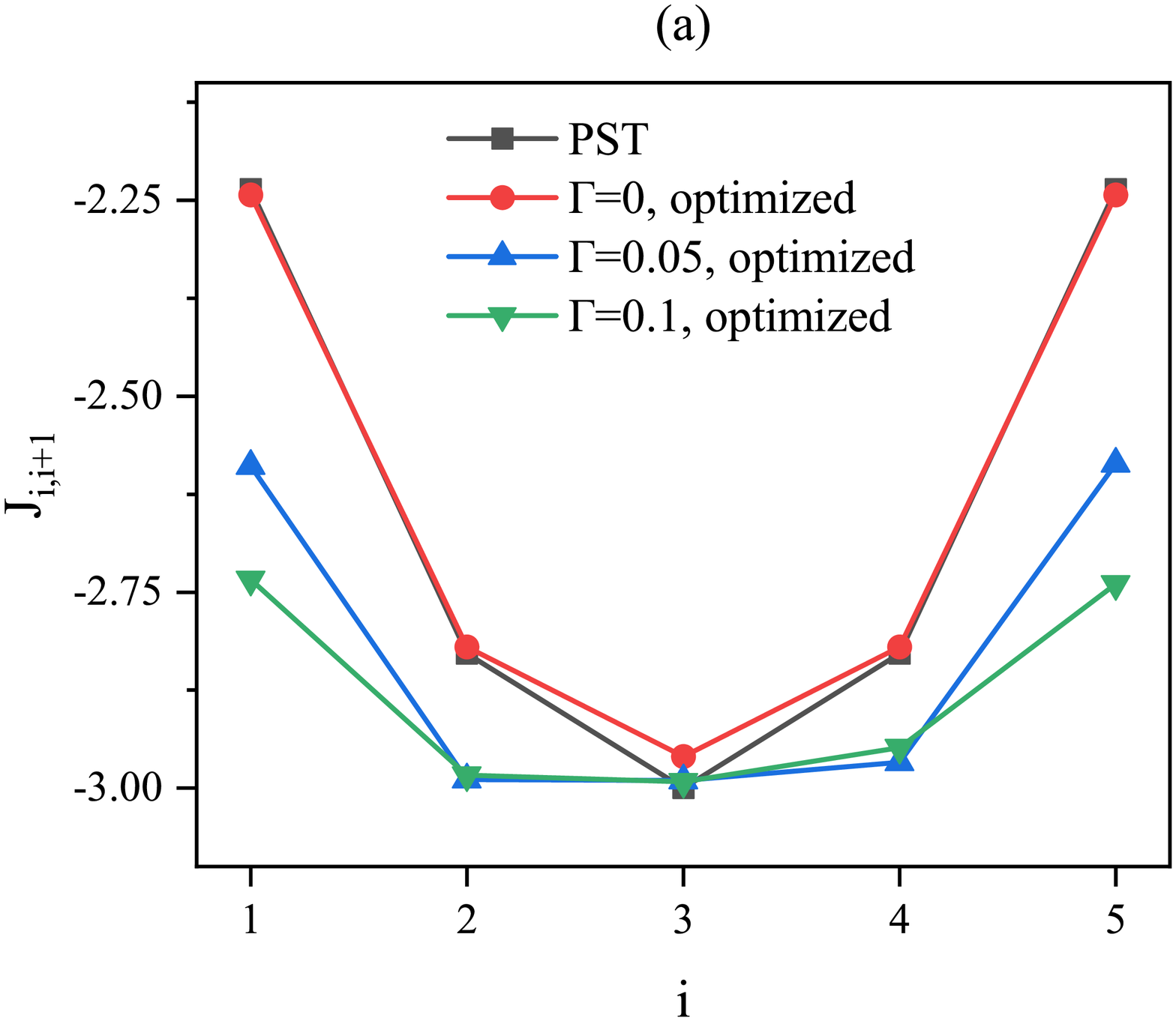}
	\includegraphics[width=\columnwidth]{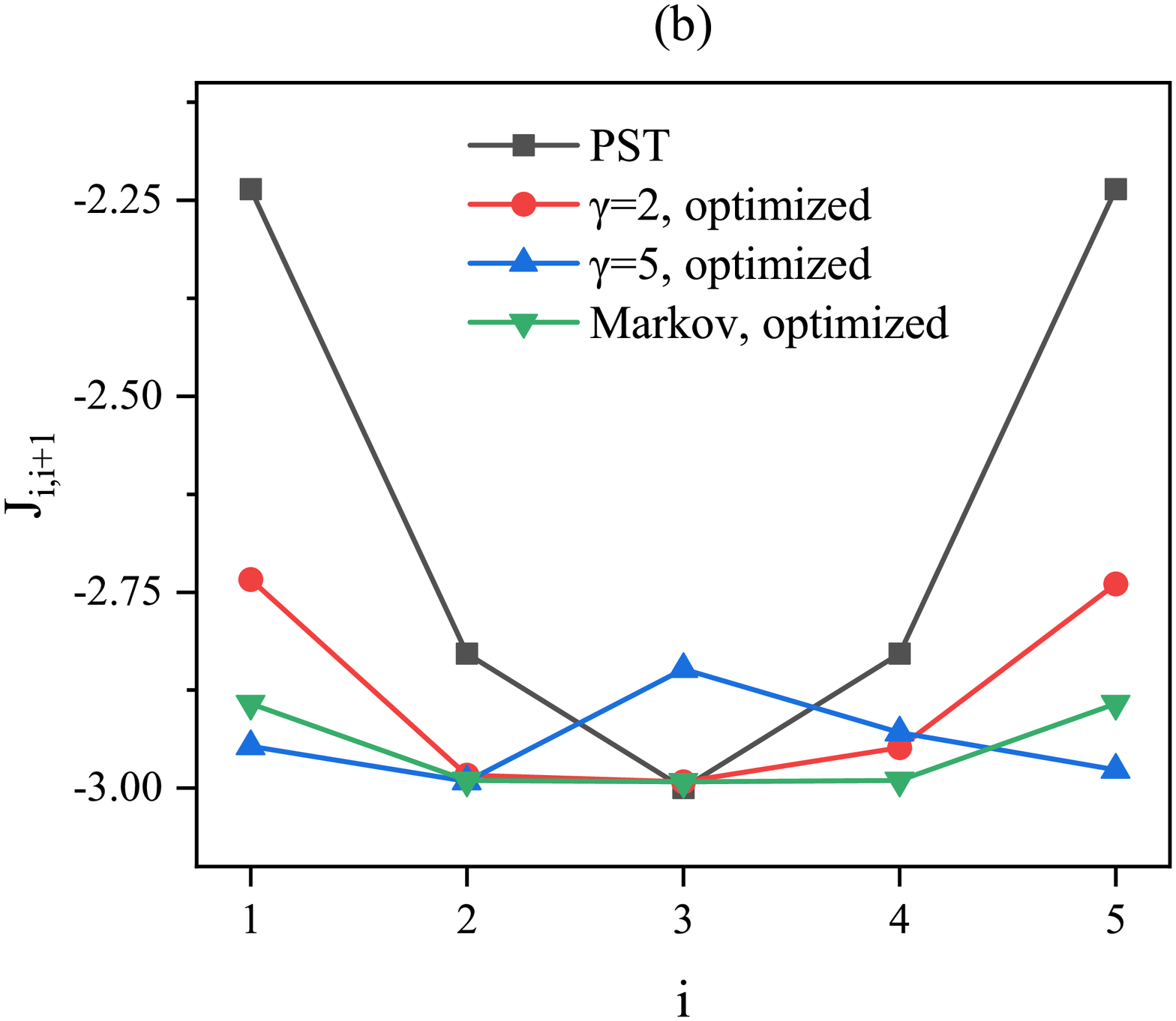}
	\includegraphics[width=\columnwidth]{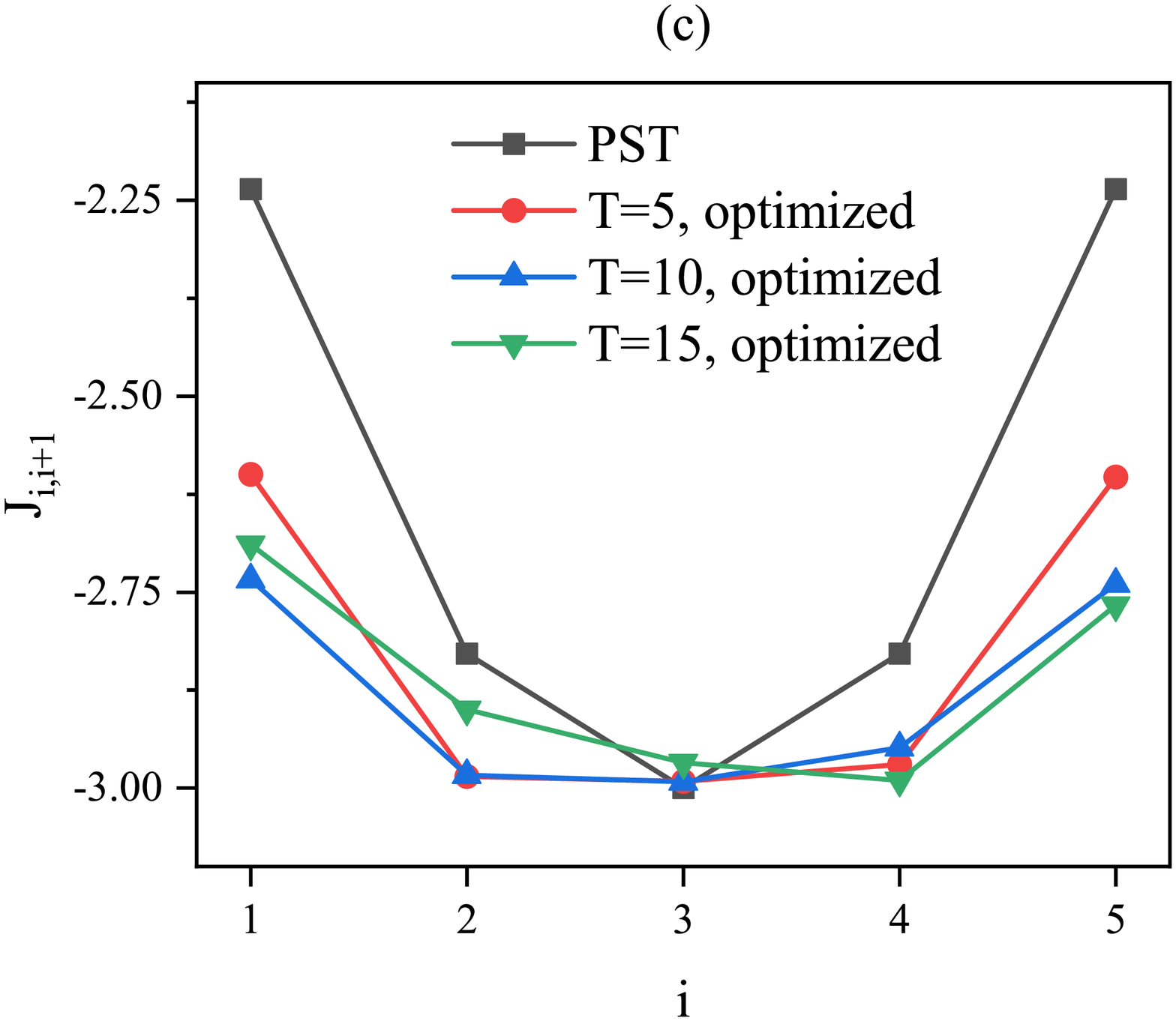}
	\caption{(Color on line) The corresponding PST and optimized couplings in Fig.~\ref{f1}.}
	\label{f2}
\end{figure}

In Fig.~\ref{f1} we plot the time evolution of the fidelity $F(t/T_{tot})$ with PST and optimized couplings for different environmental parameters. The parameters are taken as $\gamma=2$, $T=10$ (Fig.~\ref{f1}(a)), $\Gamma=0.01$, $T=10$ (Fig.~\ref{f1}(b)), and $\Gamma=0.01$, $\gamma=2$ (Fig.~\ref{f1}(c)), respectively. For a fair comparison between PST and optimized couplings, the optimized couplings are limited in $[-3,-2]$, which has the same region as PST. At first, without optimization, the exposure to environment always decreases the fidelity. A larger $\Gamma$, $\gamma$ or $T$ corresponds to a lower fidelity $F$, i.e., a stronger system-bath interaction, more Markovian or higher temperature bath will destroy the system more severely, which is in accordance with Refs.~\cite{nie2021, wang2020aest2}. This result still holds for the optimized cases. For example, in Fig.~\ref{f1}(b), the maximum fidelity $F_{max}=0.73$ is obtained for $\gamma=2$ and as $\gamma$ grows, $F_{max}$ decreases. Secondly, comparing $F_{max}$ for the PST and optimized couplings with the same environmental parameters, we find that the optimized $F_{max}$ are always higher than these in the ideal cases. In another word, using the optimized couplings, $F_{max}$ can always be enhanced in the presence of environment. This is the key observation of our paper. It is not so obvious but worth mentioning that the maximal fidelity improvement increases with increasing $\Gamma$ and $T$. That is to say, the bath destroys system more severely, the improvement is more significant. Namely, a more stronger bath provides more space for Adam to optimize. Clearly, without environment ($\Gamma=0$ in Fig.~\ref{f1}(a)), the evolution is the same for PST and optimized couplings. Thirdly, defining the arrival time $T_a$ when $F_{max}$ is achieved, with PST couplings, $T_a$ occurs at $t=\pi/4$ for different $\Gamma$ and $T$, and nearly at $t=\pi/4$ for different $\gamma$. The bath slightly affects the arrival time $T_a$ under ideal pulses. However, after optimization, $T_a$ is evidently shorter than $\pi/4$, which bears the advantage that $F_{max}$ arrives earlier and thus the accumulative detrimental effects of the environment can be partially avoided. In Fig.~\ref{f1}, $T_a$ is shorter for larger $\Gamma$, $\gamma$ and $T$. At last, even for the Markovian case (Fig.~\ref{f1}(b)), $F_{max}$ can still be enhanced by the coupling optimization. In sum, our optimized couplings via Adam algorithm can simultaneously enhance the transmission fidelity and shorten the arrival time. 

Fig.~\ref{f2} plots the corresponding PST and optimized couplings used in Fig.~\ref{f1}. The optimized coupling configuration is similar to the PST: bigger in the middle and smaller in both ends. But for a more stronger bath (bigger $\Gamma$, $\gamma$ and $T$), Adam finds a more flatter configuration. The minimum and maximum get closer. Also, the symmetry of the couplings with respect to the middle of the chain is broken due to the existence of the environments, which can be clearly seen for a strong bath ($\Gamma=0.1$ in Fig.~\ref{f2}(a), $\gamma=5$ in Fig.~\ref{f2}(b), or $T=15$ in Fig.~\ref{f2}(c)).

\subsection{Optimized pulses via Adam}

\subsubsection{QST under ideal pulse control}
\begin{figure}[!htbp]
	\centering
	\includegraphics[width=0.95\columnwidth]{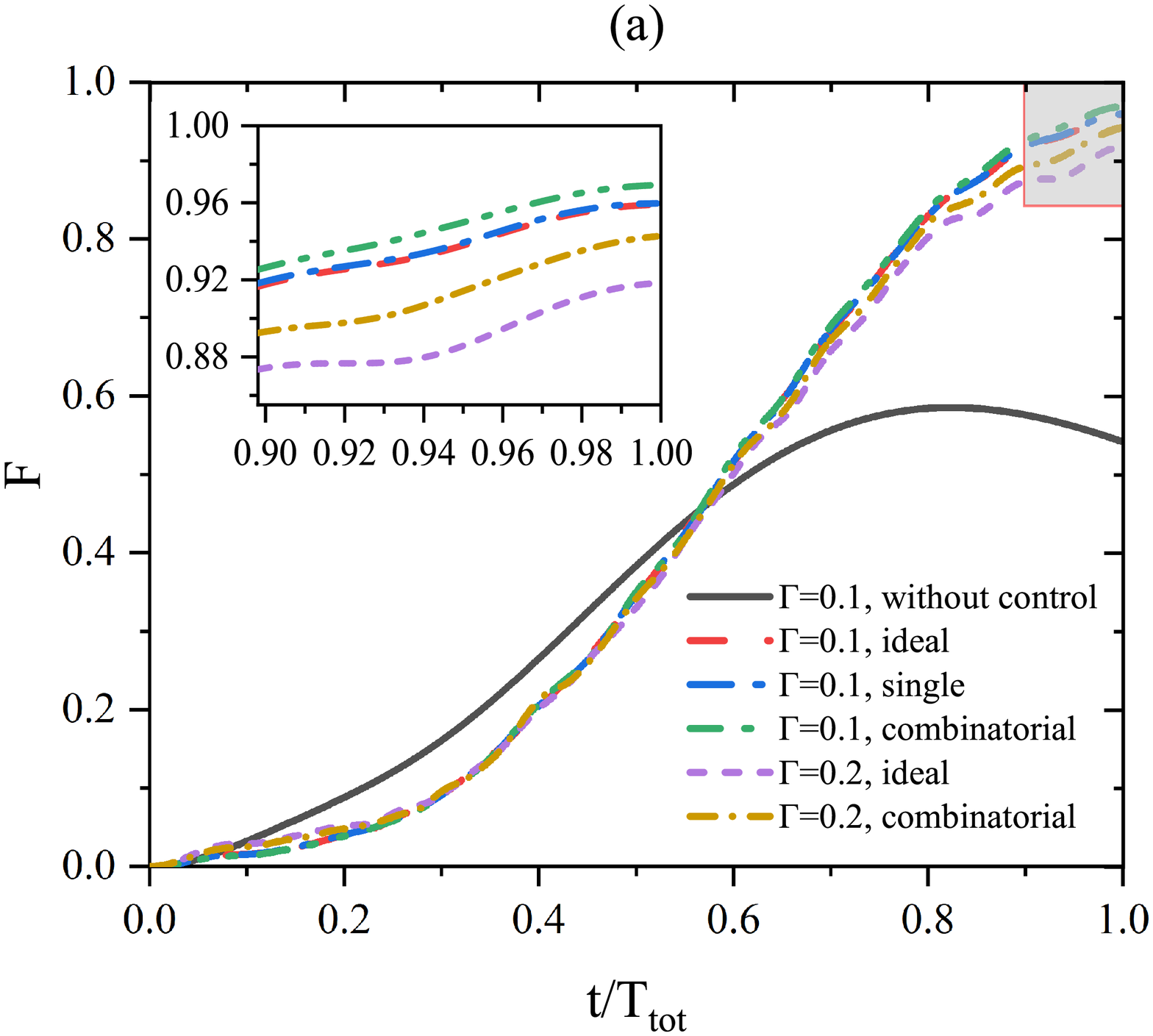}
	\includegraphics[width=0.95\columnwidth]{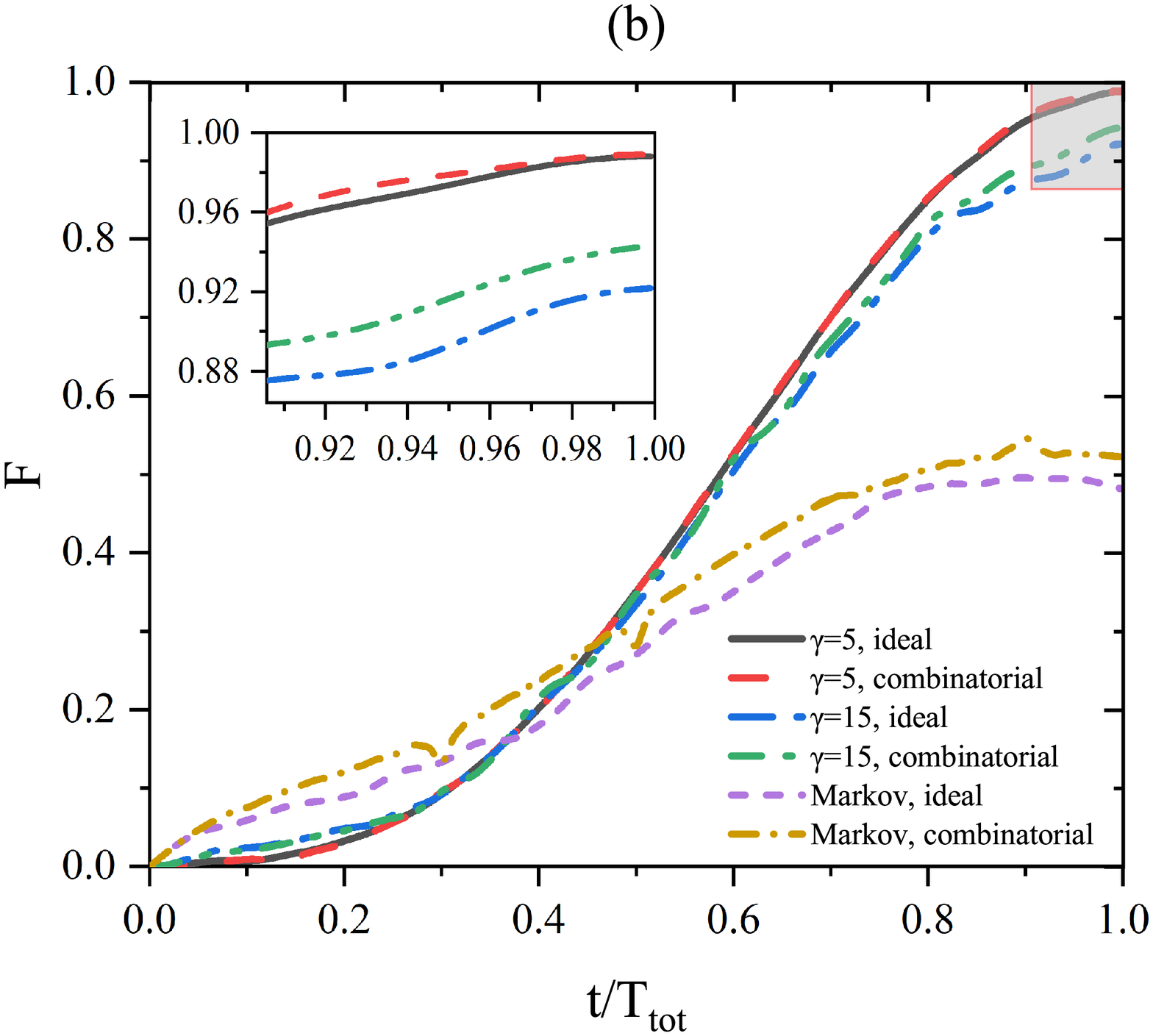}
	\includegraphics[width=0.95\columnwidth]{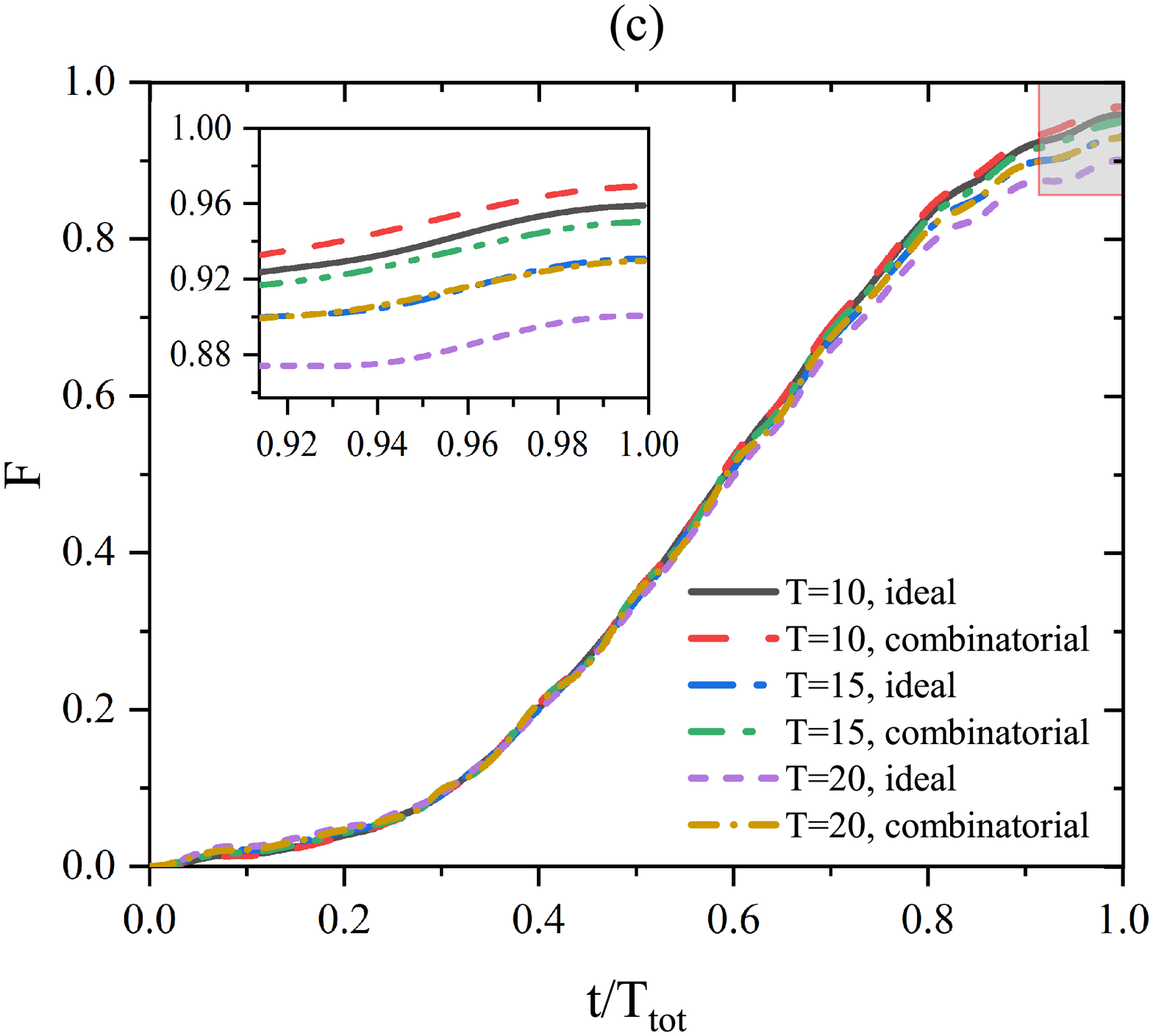}
	\caption{(Color on line) The fidelity $F$ versus the rescaled time $t/T_{tot}$ with ideal and optimized (single and combinatorial) pulses for different parameters (a) $\Gamma$, $\gamma=10$, $T=10$; (b) $\gamma$, $\Gamma=0.1$, $T=10$; (c) $T$, $\Gamma=0.1$, $\gamma=10$.}
	\label{f3}
\end{figure}

Environmental noise normally destroy the transmission fidelity and Refs.~\cite{wang2021stfinit, nie2021} have introduced an LEO approach to address this problem. The main idea of this LEO approach is to add an additional Hamiltonian $H_{LEO}$ to the system Hamiltonian $H_{s}$, ensuring the quantum system to evolve along a predefined passage. For example, if we use $H_{PST}$ to denote the Hamiltonian in Eq.~(\ref{hs}) with PST couplings, we can set $|\Psi\left(t\right)\rangle=\exp\left (-iH_{PST} t\right )|\textbf{1}\rangle$ as the evolution passage. The LEO Hamiltonian in adiabatic frame \cite{wang2018experimental} can be constructed as
\begin{equation}
	H_{LEO}=c\left (t \right )|\Psi\left(t\right)\rangle\langle\Psi\left(t\right)|,
\end{equation}
where $c(t)$ is the control function. The total Hamiltonian becomes
\begin{equation}
	H_{tot}=H_{s}+H_{LEO}.
\end{equation}
The LEO Hamiltonian can be achieved by a series of control pulses that can be divided into perturbative and nonperturbative versions. In this paper we consider the latter one whose pulse intensity and duration are finite. The pulse conditions for effective control have been theoretically deduced by P-Q partitioning technique in closed systems \cite{chen2018, zhang2019, pulsecondition}. For sine pulses $c(t)=I\sin(\omega t)$, the corresponding pulse condition is
\begin{equation}
	J_{0}(\frac{I\tau}{\pi})=0.
	\label{condition}
\end{equation}
Here $I$ and $\tau$ represent pulse intensity and half period, and $J_{0}(x)$ denotes the zero-order Bessel function of the first kind. Note that the integral of such pulses over a period is zero (i.e., zero-area condition of pulses) \cite{pulsecondition, wang2018experimental}. The control pulses such as rectangular and triangular ones have also been investigated \cite{zhang2019}.

\subsubsection{High-fidelity QST}
Although the above ideal pulse conditions are derived theoretically from closed systems, they can be applied to open ones with no guarantee of their effectiveness. In this section, we aim to design optimized pulses for certain environmental parameters with the help of Adam, and then compare their performances with ideal counterparts. In order to make fair comparisons, the optimized pulses also satisfy the zero-area condition \cite{pulsecondition, wang2018experimental}. First we design the optimized sine pulses (single pulses)
\begin{equation}
	c(t)=I(t)\sin(\omega t).
\end{equation}
Here $I(t)$ is a $P$ segment piece-wise constant function, whose $P$ values are drawn in order from the pulse amplitude sequence $\textbf{\textit{I}}=[I_{0}, I_{1}, \cdots, I_{P-1}]$ and take the equal time interval $\Delta t=T_{tot}/P$ ($\omega=2\pi/\Delta t$ and set $P=5$). Notice that the zero-area condition \cite{pulsecondition,wang2018experimental} is followed in iterative procedures as in theoretical derivation. We consider the corresponding ideal values $\textbf{\textit{I}}=[96.200, 96.200, \cdots, 96.200]$, derived from the pulse condition in Eq.~(\ref{condition}), as our initial guess. The maximal iteration number is still $k_{max}=1000$ and the number of spins $N=4$. Furthermore, the maximal intensity of the optimized pulses are not supposed to outweigh that of their ideal counterparts. Similar to Eq.~(\ref{loss1}), the loss function is accordingly defined as
\begin{equation}
	Loss(\textbf{\textit{I}})=1-F(\textbf{\textit{I}})+\lambda c_{max}.
	\label{loss2}
\end{equation}
Here $c_{max}$ is the maximum of the control function $c(t)$. In Eq.~(\ref{loss2}), there also is a competition between infidelity $1-F(\textbf{\textit{I}})$ and maximal control intensity $c_{max}$ for $Loss$, and a relaxation parameter $\lambda$ to restrict $c_{max}$ \cite{zxmpra}.

\begin{figure}[!htbp]
	\centering
	\includegraphics[width=\columnwidth]{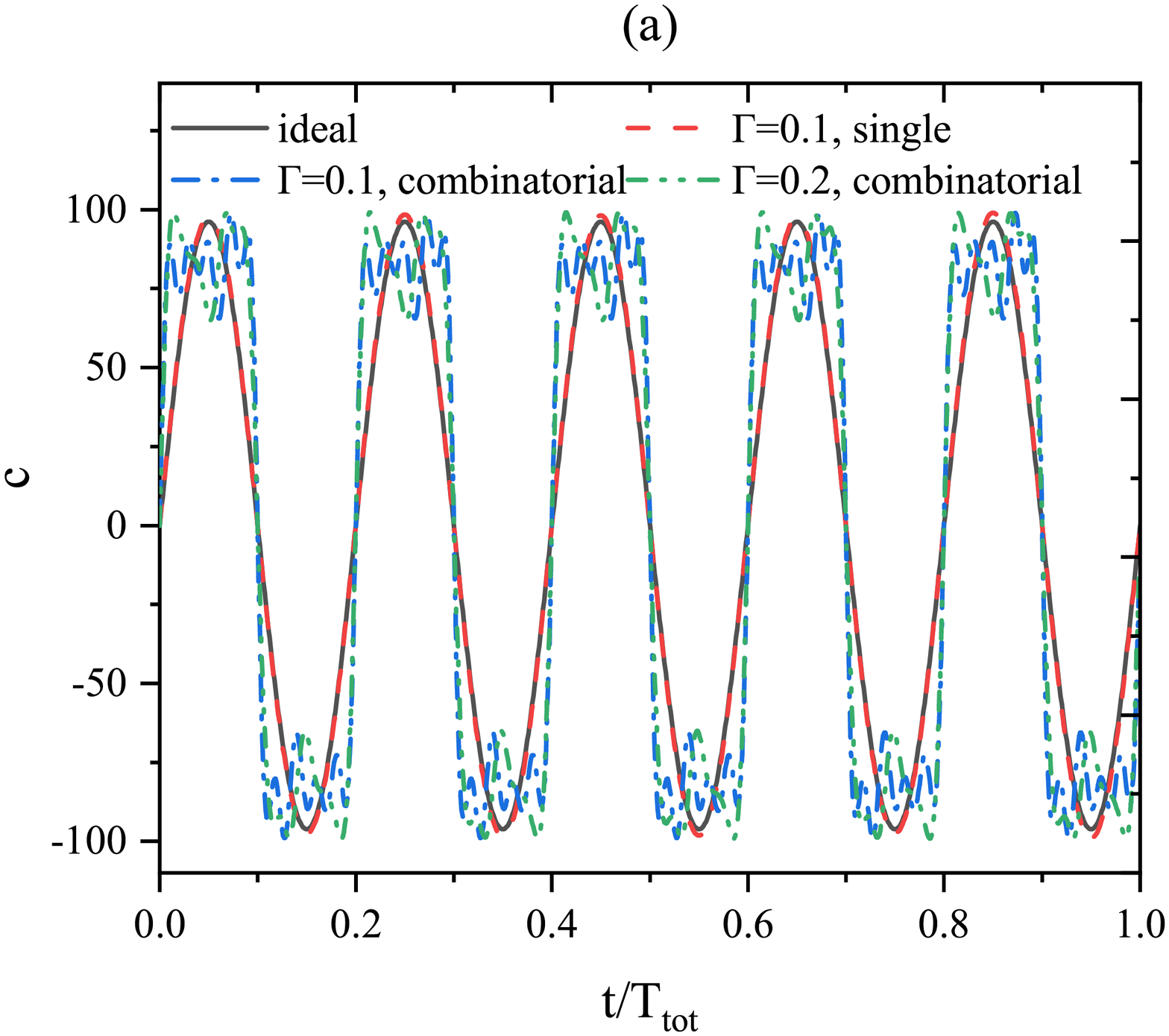}
	\includegraphics[width=\columnwidth]{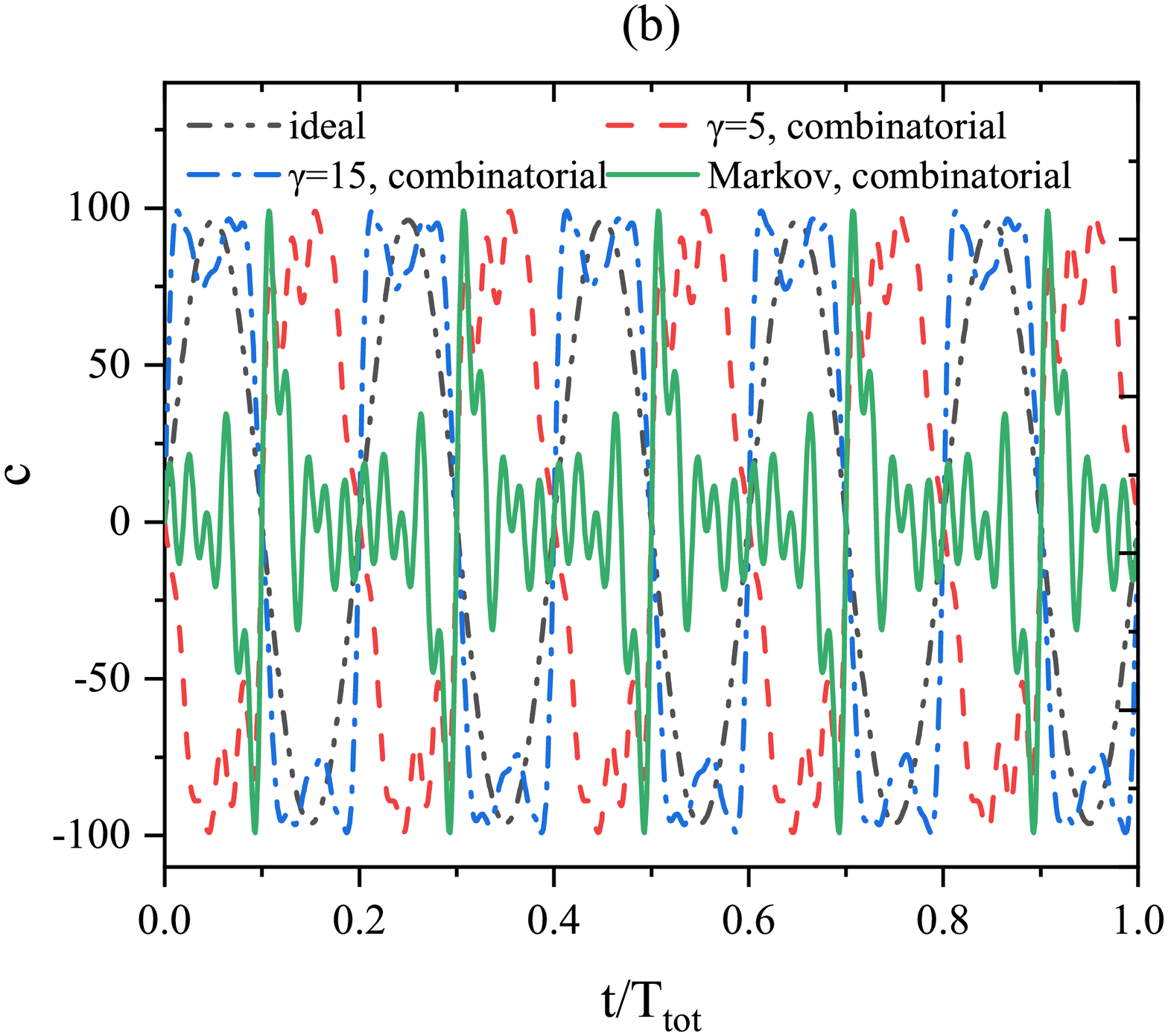}
	\includegraphics[width=\columnwidth]{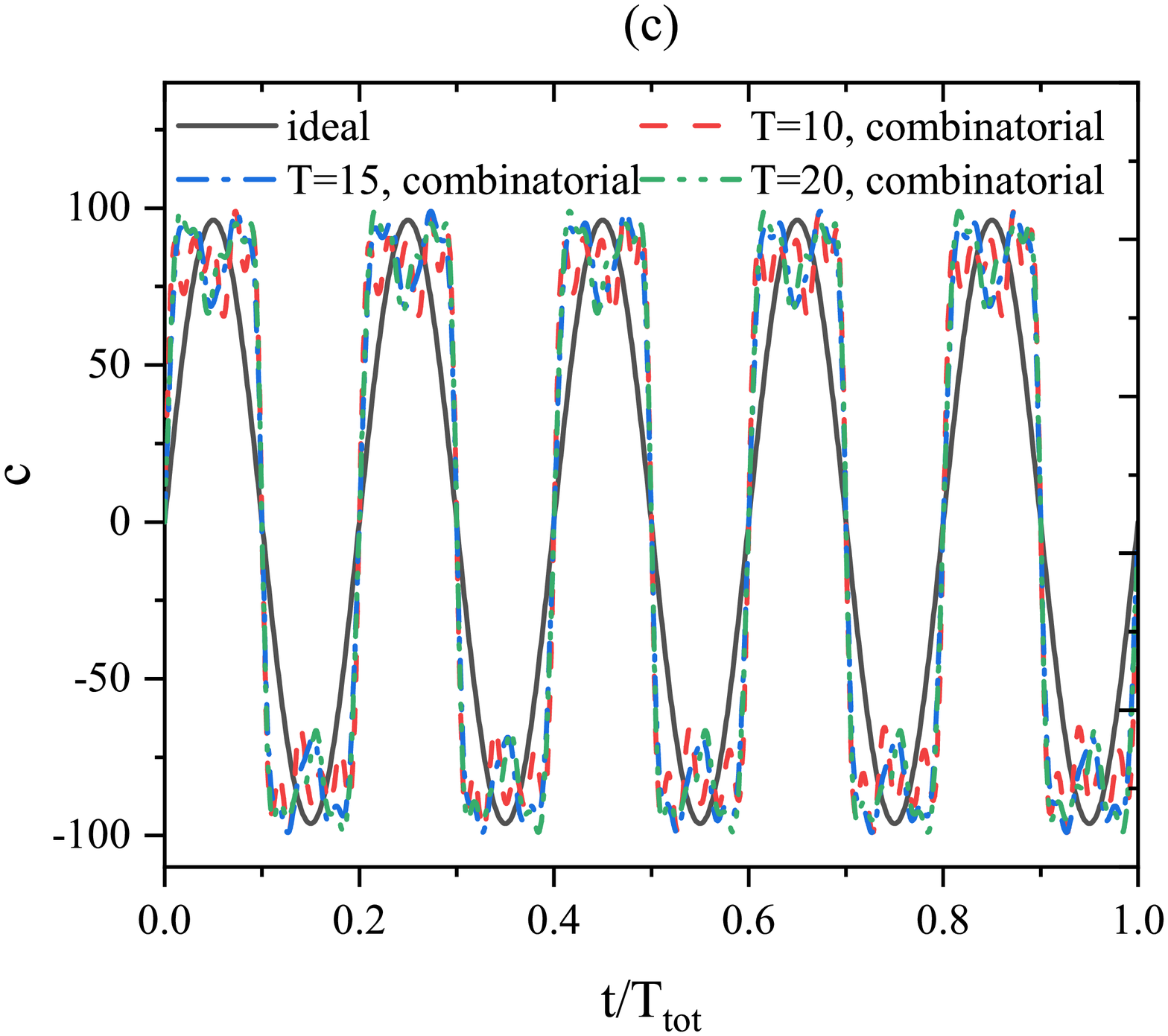}
	\caption{(Color on line) The corresponding ideal and optimized (single and combinatorial) pulses in Fig.~\ref{f3}.}
	\label{f4}
\end{figure}

In Fig.~\ref{f3}, we plot the fidelity $F$ as a function of the rescaled time $t/T_{tot}$ for different environmental parameters with ideal and optimized pulses. In Fig.~\ref{f3}(a), $\gamma=10$ and $T=10$. When $\Gamma=0.1$, the maximal fidelity $F_{max}(t)$ dramatically rockets, from $0.585$ without control to $0.958$ with ideal pulses and $0.959$ with single pulses. Note that the single pulses ultimately reach the similar fidelities as the ideal pulses can do. We then propose the combinatorial sine pulses (combinatorial pulses) to obtain a higher fidelity,
\begin{equation}
	c(t)=\sum_{i=0}^{Q-1}I_{i}\sin\left[\left(i+1\right)\omega t\right],
	\label{combinatorial}
\end{equation}
where we turn to set the control function $c(t)$ as a combination of Fourier sine components. Here $Q$ denotes the number of Fourier components and we consider $Q=10$. Notice that the zero-area condition \cite{pulsecondition, wang2018experimental} is still satisfied. Obviously, when $\Gamma=0.1$ and 0.2, combinatorial pulses overshadow the ideal and single counterparts, and there are minor but evident increases on QST fidelities. From now on, we choose to optimize combinatorial pulses alone.

In Fig.~\ref{f3}(b) and (c), we plot the influences of the parameters $\gamma$ and temperature $T$ on the fidelity. In Fig.~\ref{f3}(b), $\Gamma=0.1$ and $T=10$ while in Fig.~\ref{f3}(c), $\Gamma=0.1$ and $\gamma=10$. For all the situations, without exception, the combinatorial pulses outshine the ideal ones. Furthermore, an increasing $\Gamma$, $\gamma$ or $T$ corresponds to a decreasing fidelity $F$. But still, in a more stronger bath, the optimized pulses can make larger corrections for this fidelity deterioration. Fig.~\ref{f4} gives the profiles of corresponding ideal and optimized (single and combinatorial) pulses in Fig.~\ref{f3}. Fig.~\ref{f4}(a) shows that the single pulses are almost indistinguishable from the ideal ones. As for the combinatorial pulses, they are only similar with the ideal pulses in terms of the magnitude. From the above analysis, we conclude that the scheme of optimized control pulses can play more helpful roles than the ideal ones, especially in stronger baths.

\begin{figure}[!htbp]
	\centering
	\includegraphics[width=\columnwidth]{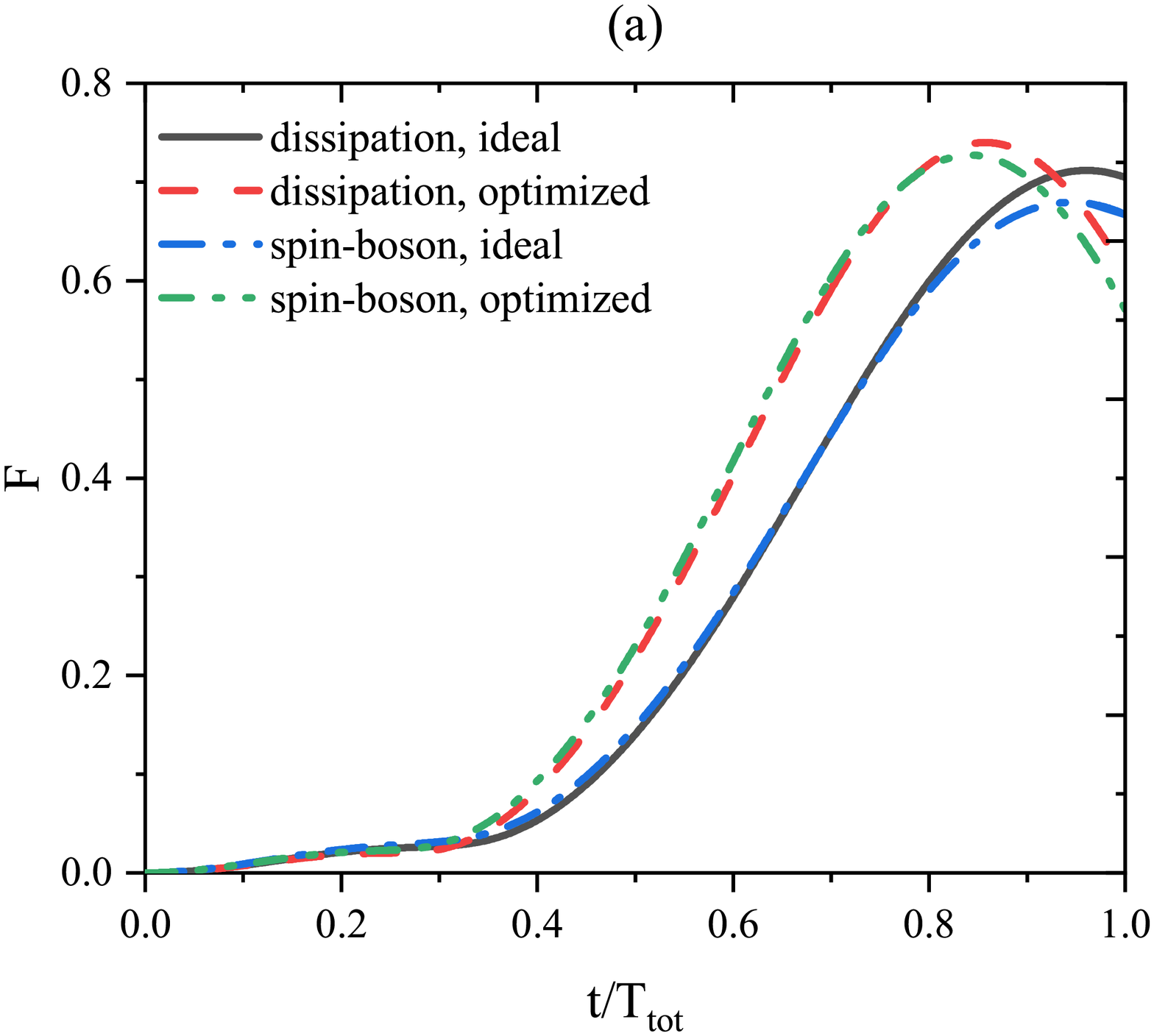}
	\includegraphics[width=\columnwidth]{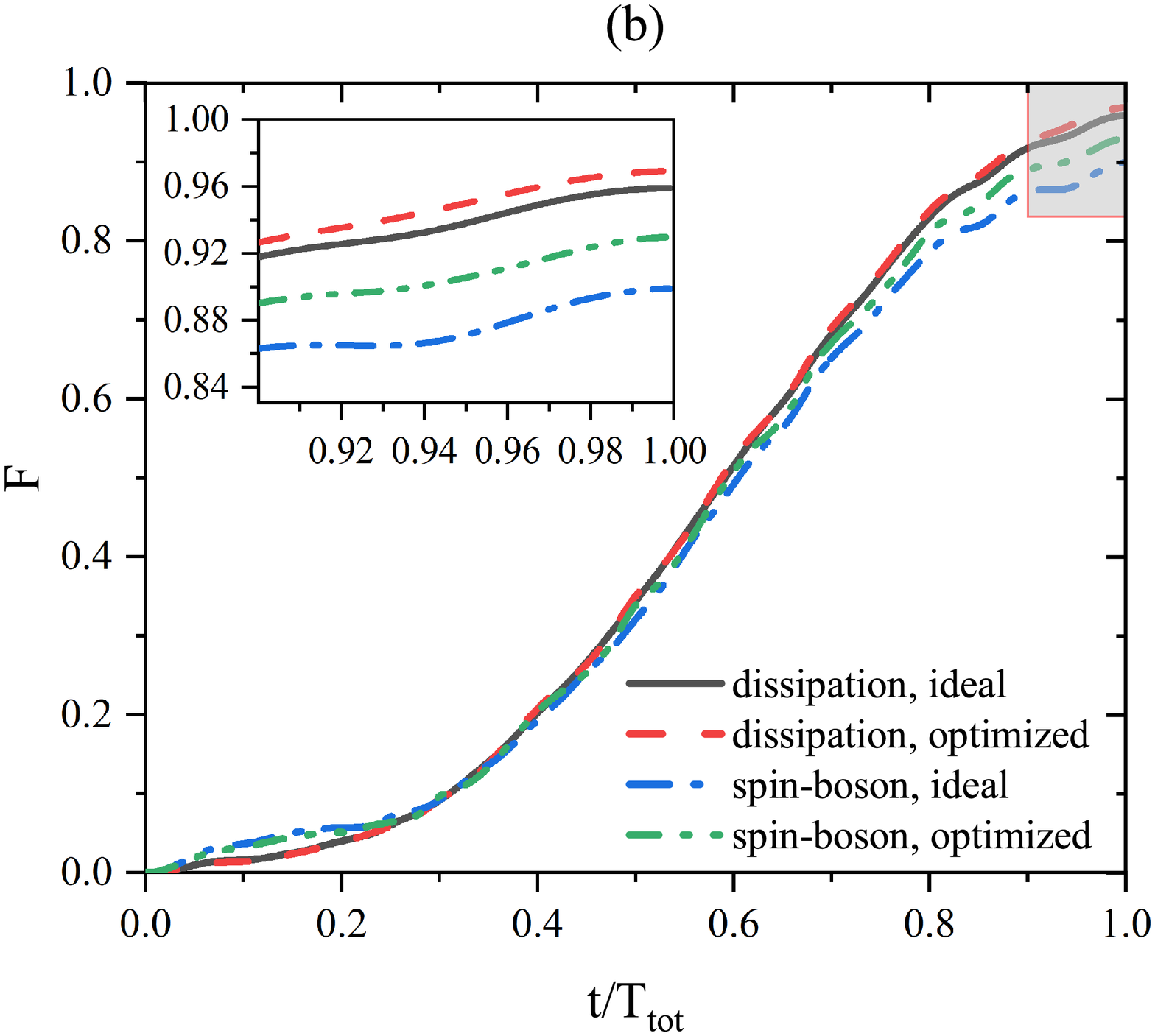}
	\caption{(Color on line) The fidelity $F$ versus the rescaled time $t/T_{tot}$ for different Lindblad operators $L=\sum_{i=1}^{N}\sigma _{i}^{x}$ and $L=\sum_{i=1}^{N}\sigma _{i}^{-}$ with (a) PST and optimized couplings; (b) ideal and optimized (combinatorial) pulses. $T=10$. $N=6$, $\Gamma=0.05$ and $\gamma=2$ in Fig.~\ref{f5}(a). $N=4$, $\Gamma=0.1$ and $\gamma=10$ in Fig.~\ref{f5}(b).}
	\label{f5}
\end{figure}

At last we consider different types of Lindblad operator $L$. We will compare the effects of $L=\sum_{i=1}^{N}\sigma _{i}^{-}$ and $L=\sum_{i=1}^{N}\sigma _{i}^{x}$, and the latter corresponds to the spin-boson interaction. We do not consider the dephasing ($L=\sum_{i=1}^{N}\sigma _{i}^{z}$) because $\left [ L,\rho_{s}\overline{O}^{\dag} \right ]=\left [ L^{\dag},\overline{O}\rho_{s} \right ]=0$. Therefore the bath only randomly changes the global phase of system \cite{wang2020aest}. In Fig.~\ref{f5} we plot the cases with different Lindblad operators. Fig.~\ref{f5}(a) shows that the fidelity obtained by the optimized couplings exceeds the PST ones whatever the Lindblad operator $L$ is. The parameters are taken as $N=6$, $\Gamma=0.05$, $\gamma=2$ and $T=10$. Fig.~\ref{f5}(b) demonstrates the implications of $L$ on performances of optimized pulses. Again optimized pulses show their advantage over the ideal counterparts on reducing the effects of environmental noise. We take $N=4$, $\Gamma=0.1$, $\gamma=10$ and $T=10$ in the simulation.

\section{Conclusions}
QST is one of the basic tasks in quantum computation. PST and almost exact QST through a spin chain can be realized for PST couplings and LEO control, respectively. However, theses conditions are derived theoretically from the ideally closed systems and thus their effectiveness are lost when they are applied to an open system, i.e., being coupled to a heat bath results in their dissipation dynamics. In this paper, we take a one-dimensional $XY$ spin chain with nearest-neighbor couplings as an example and introduce a well-developed optimization algorithm, Adam, to seek for the optimized couplings and control pulses in the presence of environment. By minimizing a predefined loss function, high-fidelity transmission is obtained for both schemes. In addition, we discuss the effects of system-bath coupling strength $\Gamma$, environmental non-Markovianity parameter $\gamma$ and temperature $T$ on our schemes. Although the fidelity $F$ decreases with anyone of these parameters increasing, our optimized schemes perform better, especially for a stronger bath. Our work shows that the Adam algorithm is a powerful tool to search the optimized parameters in open quantum systems, which are important in performing quantum information processing tasks.

\section*{Acknowlegment}
This paper is based upon work supported by the Natural Science Foundation of Shandong Province (Grants No. ZR2021LLZ004).

\bibliography{References_library.bib}

\end{document}